\documentclass[5p,twocolumn]{elsarticle}
\usepackage{mathrsfs}
\usepackage[dvipsnames,usenames]{color}
\usepackage{amsmath}
\usepackage{amssymb}
\usepackage{graphicx}
\usepackage{times}
\usepackage{graphics,color}
\usepackage{hyperref}
\newtheorem{theorem}{Theorem}
\newtheorem{lemma}{Lemma}

\newtheorem{corollary}{Corollary}
\newtheorem{definition}{Definition}
\newtheorem{remark}{Remark}
\newcommand{\onetom}{1,\cdots,m}

\newcommand{\onetoN}{1,\cdots,N}

\def\tp{\tilde{P}}

\begin{document}

\begin{frontmatter}
\title{Pinning dynamic systems of networks with Markovian switching couplings and controller-node set}
\author[fdu]{Yujuan Han}
\ead{09210180039@fudan.edu.cn}
\author[fdu,fdub,max]{Wenlian Lu\corref{cor1}}
\ead{wenlian@fudan.edu.cn}
\author[fdu]{Zhe Li}
\ead{11210180026@fudan.edu.cn}
\author[fdu,fduc]{Tianping Chen}
\ead{tchen@fuan.edu.cn}
\cortext[cor1]{Corresponding author.}
\address[fdu]{School of Mathematical Sciences, Fudan University, Shanghai 200433, China}
\address[fdub]{Centre for Computational Systems Biology, Fudan University, Shanghai 200433, China}
\address[max]{Department of Computer Science, The University of Warwick, Coventry CV4 7AL, United Kingdom}
\address[fduc]{School of Computer Science, Fudan University, Shanghai 200433, China}
\fntext[fn1]{This work is jointly supported by the National Key Basic Research and
Development Program (No. 2010CB731403), the National Natural Sciences
Foundation of China under Grant Nos. 61273211, 60974015, and 61273309, the Marie Curie International Incoming Fellowship
from the European Commission (FP7-PEOPLE-2011-IIF-302421), and also the Laboratory of Mathematics for Nonlinear Science, Fudan University.}

\begin{abstract}
In this paper, we study pinning control problem of coupled dynamical
systems with stochastically switching couplings and stochastically selected
controller-node set. Here, the coupling matrices and the controller-node
sets change with time, induced by a continuous-time Markovian chain. By
constructing Lyapunov functions, we establish tractable sufficient
conditions for exponentially stability of the coupled system. Two scenarios are
considered here.  First, we prove
that if each subsystem in the switching system, i.e. with the fixed
coupling, can be stabilized  by the fixed pinning controller-node set, and
in addition, the Markovian switching is sufficiently slow, then the
time-varying dynamical system is stabilized. Second, in
particular, for the problem of spatial pinning control of network with
mobile agents, we conclude that if the system with the average coupling and
pinning gains can be stabilized and the switching is sufficiently fast, the
time-varying system is stabilized. Two numerical examples are provided to
demonstrate the validity of these theoretical results, including a switching
dynamical system between several stable sub-systems, and a dynamical system
with mobile nodes and spatial pinning control towards the nodes when these
nodes are being in a pre-designed region.
\end{abstract}
\end{frontmatter}

\section{Introduction}

Control and synchronization of large-scale dynamical systems have attracted
wide interests over the past decades since the new discoveries of
small-world and scale-free features \cite{Watts,Barabasi}. When the network
cannot be stabilized by itself, many control strategies are taken into
account to force the network to be stable. Among these control strategies,
pinning control is brilliant because
it is easily realizable by controlling a partial of
the nodes instead of all nodes in the network.

The general idea behind pinning control is to apply some local feedback
controllers only to a fraction of nodes and the rest of nodes can be
propagated through the coupling among nodes. In most existing papers, authors
considered two different pinning strategies: randomly pinning and selective
pinning based on the connectivity degrees. \cite{Wang02,Li04} concluded
that pinning of the most highly connected nodes performance better
stabilizability than the randomly pinning scheme. \cite{Chen07}-\cite{Yu09}
used Lyapunov functions, master stability functions and algebraic
properties of the coupling matrix to study pinning control of complex
network. \cite{Lu10}-\cite{Liuhy} gave conditions based on the graph
topology of the network guaranteeing stability. In these papers, networks with time-invariant links were considered.
\cite{Yu13} studied the global pinning synchronization of
Lorenz-type dynamical networks with both fixed and switching topologies.
It was proved that if each possible network topology contains a directed spanning tree and the
dwell time of switching is larger than a positive threshold, global pinning synchronisation in switching networks with a suitable coupling strength can be guaranteed. However,
pinning dynamic systems of networks with stochastically switching topologies
is absent.

Noting that stability and controllability of stochastic differential equations have been well
studied these years \cite{Arnold}-\cite{Kolmanovskii}, in which
differential dynamical systems with Markovian switching have received lots
of research interests \cite{Ji}-\cite{Mao06}. In these papers, the stability of
linear, semi-linear and nonlinear equations with  Markovian switching were
studied and some sufficient criteria were established based on Lyapunov
functions and linear matrix inequalities. Hence,
in this paper, we investigate the pinning control of networks with Markovian switching topologies.

However, for a network of mobile agents, it is difficult to implement a pinning
control strategy because the selected controllers are moving.
\cite{Wang10}-\cite{Yu10} investigated the flocking control
via selecting some fixed nodes to be pinned prior.
To avoid this, \cite{Frasca} introduced the concept of spatial
pinning control for network via applying controllers only on nodes (agents)
which enter a given area, called control region. In its setup, agents
move following random walking in a planar space and interact according to a
time-varying $r$-disk proximity graph. In the this paper, we considered
a more general case, the moving motion of each agent follows a Markov
process.

Inspired by these works,
we investigate the pinning control of a complex network with
Markovian switching couplings to an arbitrary trajectory of the uncoupled node
system by Markovian selected controller-node set. We derive analytical conditions for the stability of the system at the homogeneous trajectory. Based on the
switching speed of the couplings and controller-node sets, two scenarios are
considered here.
Firstly, if each subsystem in the switching system with fixed couplings
can be stabilized by fixed pinning nodes and the Markovian switching is sufficiently slow, the system can be stabilized. Secondly, if the system with average coupling matrix and pinning gains can be stabilized and the Markovian switching is sufficiently fast,
the system can be stabilized.
As an application, we also study the spatial pinning control problem of network with mobile agents.

This paper is organized as follows. Section 2 presents some
definitions and some notations required in this paper. Section 3
investigates the pinning control problem of network with Markovian
switching coupling. Section 4 investigates the fast switching networks and as an application, the spatial pinning control of
network of mobile agents is considered. Simulations are given in section 5 to verify the
theoretical results. Finally, conclusions are drawn in section 6.

\section{Preliminaries}

At first, we introduce some notations needed throughout this paper. For a
matrix $A$, denote $a_{ij}$ the elements of $A$ on the $i$th row and $j$th
column. $A^{\top}$ denotes the transpose of $A$. $A^{s}=(A+A^{\top})/2$
denotes the symmetry part of a square matrix $A$. Denote by $A>0(\geq 0)$
that $A$ is a positive(semi-positive) definite and so is with $<0$ and
$\leq 0$. $I_{m}$ denotes the identity matrix with dimension $m$. For a symmetric
square matrix $A$, denote $\lambda_{M}(A)$ and $\lambda_{m}(A)$ the largest
and smallest eigenvalues of $A$ respectively, $\lambda_i(A)$ the i-th largest eigenvalue of $A$.
$\|z\|$ denotes a vector norm of a vector $z$, $\|A\|$ denotes the matrix norm of $A$ induced by the vector norm $\|\cdot\|$ and denote $\|A\|_{\infty}=\max_{i,j}|a_{ij}|$. In particular, without special notes,
$\|z\|=\|z\|_{2}=\sqrt{\sum_{i}|z_{i}|^{2}}$, the $2$-norm. The symbol
$\otimes$ is the Kronecker product.

Linearly coupled ordinary differential equations (LCODEs) are used to
describe coupling dynamical systems, which can be
described as follows:
\begin{eqnarray}
\nonumber\dot{x}^{i}=f(x^{i},t)+\kappa\sum_{j=1,j\neq i}^{m}l_{ij}(t)\Gamma [x^{j}(t)-x^{i}(t)],&&\\
~i=\onetom.&&\label{LCODE}
\end{eqnarray}
where $x^{i}(t)=[x^{i}_{1}(t),\cdots,x^{i}_{n}(t)]\in \mathbb{R}^{n}$ is
the state variable of the $i$th node, $t\in [0,+\infty)$ is the continuous
time, $f$: $\mathbb{R}^{n}\times \mathbb{R}^{+}\mapsto \mathbb{R}^{n}$ is
the dynamics of the uncoupled nodes, $\kappa$ is the coupling strength,
$l_{ij}(t)$ denotes the coupling coefficient from agent $j$ to agent $i$ at
time $t$, $\Gamma=[\gamma_{kl}]_{k,l=1}^{n}\in \mathbb{R}^{n\times n}$ denotes the
inner connection matrix with $\gamma_{kl}\neq 0$ if two agents are
connected by their $k$th and $l$th state component respectively. For $i\neq
j$, let $l_{ii}(t)=-\sum_{j=1,j\neq i}^{n}l_{ij}(t)$, then the coupling
matrix $L(t)=[l_{ij}(t)]$ is a Metzler matrix with zero row sums at time
$t$.

Suppose $\sigma_{t}$ is a homogeneous continuous Markov chain with a finite
state space $\mathbb{S}=\{1,2,\cdots,N\}$ and its generator
$Q=[q_{ij}]_{N\times N}$ is given by
\begin{eqnarray*}
  \mathbb{P}\{\sigma_{t+\Delta}=j|\sigma_{t}=i\}
  =\left\{\begin{array}{lr}q_{ij}\Delta +o(\Delta), & i\ne j,\\
  1+q_{ii}\Delta+o(\Delta),& i=j,
  \end{array}
  \right.
\end{eqnarray*}
where $\Delta>0$, $\lim_{\Delta \to 0}(o(\Delta)/\Delta)=0$,
$p_{ij}=-\frac{q_{ij}}{q_{ii}}>0$ is the transition probability from $i$ to
$j$ if $j\ne i$, while $q_{ii}=-\sum_{j=1,j\ne i}^{N}q_{ij}$.

Denote $P=[p_{ij}]$ the transition matrix of this Markov process. Let
$\Delta_{k},k=0,1,\cdots$ be the successive sojourn time between jumps.
The sojourn time in state $i$ is exponentially distributed with parameter
$q_{i}$.

Denote $\pi(k)=[\pi_1(k),\cdots,\pi_N(k)]$ the state distribution of the
process at the $k-$th switching. From the Chapman-Kolmogorov equation
\cite{Bremaud}, the $k-$th time distribution can be expressed in terms of
the initial state distribution and the transition matrix, that is
$\pi(k)=\pi(0)P^{k}$. Suppose the Markov chain $\sigma_t$ is ergodic, then
from \cite{Billingsley}, we have that $P$ is a primitive matrix and there
exists a state distribution $\bar{\pi}=[\bar{\pi}_1,\cdots,\bar{\pi}_N]$
with positive entries satisfies $\bar{\pi}=\bar{\pi} P$. From \cite{Horn},
we obtain that there exists positive numbers $M$ and $\zeta<1$ such that
\begin{eqnarray}\label{distribution}
|\pi_{j}(k)-\bar{\pi}_{j}|<M\zeta^k.
\end{eqnarray}

Pick
$\pi_j=\frac{\bar{\pi}_j/q_j}{\sum_{i=1}^N\bar{\pi}_i/q_i},j=1,\cdots,N$
and $\pi=[\pi_1,\cdots,\pi_N]$. Then, we have $\pi Q=0$. We call $\pi$ the
\emph{invariant distribution} of Markov process $\sigma_t$.

Throughout the paper, we assume $f(x^i,t)$ belongs to the following function
class QUAD $(G,\alpha\Gamma,\beta)$.

\begin{definition}

Function class QUAD $(G,\alpha\Gamma,\beta)$: let $G$ be an $n\times n$ positive definite matrix and $\Gamma$ be an $n\times n$ matrix. QUAD$ (G,\alpha\Gamma,\beta)$ denotes a class of
continuous functions $f(\xi,t):\mathbb{R}^{n}\times[0,+\infty)\mapsto
\mathbb{R}^{n}$ satisfying
\begin{eqnarray*}\label{f}
&&(\xi-\zeta)^{\top}G[f(\xi,t)-f(\zeta,t)-\alpha \Gamma (\xi-\zeta)]\\
&&\le -\beta(\xi-\zeta)^{\top}(\xi-\zeta),
\end{eqnarray*}
holds for some $\beta>0$ and all $\xi, \zeta \in \mathbb{R}^n$.
\end{definition}

We say $f(\cdot,t)$ is globally Lipschitz if $\|f(\xi,t)-f(\zeta,t)\|\leq Lf\|\xi-\zeta\|$ holds for
some $Lf>0$ and all $\xi, \zeta \in \mathbb{R}^n$.

QUAD condition defines a class of functions arising in stability by quadratic type of Lyapunov function and it is also known as $V$-decreasing in some contexts \cite{Lu10}. It has a strong connection to other function classes, such as Lipschtiz condition and contraction. As mentioned in \cite{Luchaos},  QUAD condition is weaker than Lipschtiz condition.
since any globally Lipschitz continuous function can be QUAD for sufficiently large
$\alpha$ and $G=\Gamma=I_n$.

In this paper, we consider the pinning controlled network as follows:
\begin{eqnarray}
\nonumber&\dot{x}^{i}(t)=f(x^{i}(t),t)+\kappa\sum_{j=1}^{m}l_{ij}(\sigma_{t})\Gamma x^{j}(t)\\
&+\kappa \epsilon
c_{i}(\sigma_{t})\Gamma(s(t)-x^{i}(t)),~i=\onetom.\label{syst1}
\end{eqnarray}
where $\epsilon>0$ is the feedback control gain, $c_{i}(\tau)$ is a
variable that takes values $0, 1$, $\sigma_{t}$ is a homogeneous Markov
chain proposed in section 2, $s(t)$ satisfying $\dot{s}(t)=f(s(t),t)$ is
the target trajectory of node dynamics. Here, $s(t)$ may be an equilibrium
point, a periodic orbit, or even a chaotic orbit. The initial value
$x(0)=x_0$ is chosen randomly and independent of the other random
variables. Denote $L(\sigma_t)=[l_{ij}(\sigma_t)]$ and
$C(\sigma_t)=diag\{c_1(\sigma_t),\cdots,c_m(\sigma_t)\}$.

We define the pinning control problem as a synchronizing all the states
of the nodes in the dynamical network to an arbitrary trajectory $s(t)$ of
the uncoupled dynamical system.
Therefore, in the following, we study the stability of $s(t)$ in system (\ref{syst1}).
\begin{definition}
System (\ref{syst1}) is said to be exponentially stable at $s(t)$ in mean
square sense, if there exists constants $\delta>0$ and $M>0$, such that
\begin{eqnarray}
\mathbb{E}\bigg[\|x^{i}(t)-s(t)\|^2\bigg]\leq M e^{-\delta t}
\end{eqnarray}
holds for all $t>0$ and any $i=1,\cdots,m$.
\end{definition}
\begin{definition}
The system (\ref{syst1}) is stable almost surely if
\begin{eqnarray}
\mathbb{P}\bigg[\lim_{t\to+\infty}\|x^{i}(t)-s(t)\|=0\bigg]=1
\end{eqnarray}
holds for all $i=1,\cdots,m$.
\end{definition}

\section{Pinning time-varying networks}
In this section, we suppose $\sigma_t$ is with a finite state space
$\mathbb{S}=\{1,\cdots,N\}$.
\begin{theorem}\label{slow}
Suppose $f(x^i,t)\in QUAD(G,\alpha\Gamma,\beta)$ and there exist diagonal
positive definite matrices $P_{i}, i=\onetoN$ such that
\begin{eqnarray}\label{slow condition}
\nonumber&&\{P_{i}[\alpha I_{m}+\kappa L(i)-\kappa\epsilon C(i)]\otimes G\Gamma\}^s\\
&&+\sum_{j=1}^{m}q_{ij}P_{j}\otimes G\leq 0,~\rm{for~ all}~ i\in \mathbb{S},\label{c}
\end{eqnarray}
then the system (\ref{syst1}) is exponentially stable at the homogeneous trajectory in mean square sense.
\end{theorem}
{\em Proof.~} Let
$x=[{x^{1}}^{\top},\cdots,{x^{m}}^{\top}]^{\top}$,$\hat{s}=[s^{\top},\cdots,s^{\top}]^{\top}$,
$y=x-\hat{s}$, $F(x)=[f(x^{1})^{\top},\cdots,f(x^{m})^{\top}]^{\top}$,
$C(\sigma_{t})=diag\{c_{1}(\sigma_{t}),\cdots,c_{m}(\sigma_{t})\}$ and
${\tp}_{\sigma_{t}}=P_{\sigma_{t}}\otimes G$, then we define
$V(y,t,\sigma_{t})=\displaystyle\frac{1}{2}y^{\top}{\tp}_{\sigma_{t}}y$.
The joint process $\{(y(t),\sigma_{t}):t>0\}$ is a strong Markov process
and the infinitesimal generator of the process is:
\begin{eqnarray*}
\mathcal{L}=Q+diag\{y^{\top}P^{\top}(1)\frac{\partial}{\partial y},
\cdots,y^{\top}P^{\top}(m)\frac{\partial}{\partial y}\}
\end{eqnarray*}
Then, we have
\begin{align}\label{LV}
\mathcal{L}V(y,t,i)&=&\sum_{j=1}^{N}q_{ij}V(y,t,j)+(\frac{\partial
V(y,t,i)}{\partial y})^{\top}\dot{y},
\end{align}

Let $0<\delta\leq
\frac{2\beta\min_{i,j}(P_{j})_{ii}}{\max_{i}\{\lambda_{M}(P_{i})\}}$. From
the Dynkin Formula \cite{Mao06}, we have
\begin{align*}
&\mathbb{E}e^{\delta t}V(y,t,\sigma_{t})\\
=&V(y_0,0,\sigma_{0})+\delta\mathbb{E}\int_{0}^{t}e^{\delta \tau}V(y,\tau,\sigma_{\tau})d\tau\\
&+\mathbb{E}\int_{0}^{t}e^{\delta \tau}\mathcal{L}V(y,\tau,\sigma_{\tau})d\tau\\
\le & V(y_0,0,\sigma_{0})+\delta\mathbb{E}\int_{0}^{t}e^{\delta \tau}V(y,\tau,\sigma_{\tau})d\tau \\
&-\min_{i,j}(P_{j})_{ii}\mathbb{E}\int_{0}^{t}e^{\delta \tau}\beta y^{\top}yd\tau\\
&+\mathbb{E}\int_{0}^{t}e^{\delta \tau}y^{\top}\{P_{\sigma_{\tau}}[\alpha
I_{m}
+\kappa L(\sigma_{\tau})-\kappa\epsilon C(\sigma_{\tau})]\\
& \otimes G\Gamma
+\sum_{j=1}^{m}q_{\sigma_{\tau}j}P_{j}\otimes G\}^s yd\tau\\
\le & V(y_0,0,\sigma_{0})+\delta\mathbb{E}\int_{0}^{t}e^{\delta \tau}V(y,\tau,\sigma_{\tau})d\tau \\
&-\min_{i,j}(P_{j})_{ii}\mathbb{E}\int_{0}^{t}e^{\delta \tau}\beta y^{\top}yd\tau\\
\le & V(y_0,0,\sigma_{0})+\mathbb{E}\int_{0}^{t}e^{\delta \tau}[-\min_{i,j}(P_{j})_{ii}\beta\\
&+\frac{1}{2}\delta\max_{i}\{{\lambda_{M}({\tp}_{i})}\}]y^{\top}yd\tau\\
\le & V(y_0,0,\sigma_{0})
\end{align*}
The first inequality is because of (\ref{LV}), $f(x,t)\in
QUAD(G,\alpha\Gamma,\beta)$ and the second inequality is due to the
assumption (\ref{slow condition}). Since for each $j\in \mathbb{S}$, we
have
\begin{eqnarray*}
 \min_{i}\lambda_{m}({\tp}_{i})y^{\top}y\le y^{\top}{\tp}_{j}y\le
  \max_{i}\lambda_{M}({\tp}_{i})y^{\top}y
\end{eqnarray*}

Thus,
\begin{eqnarray*}
&&\mathbb{E}e^{\delta t}\|x^{j}(t)-s(t)\|^{2}\\
&\le& \frac{1}{\min_{i}\lambda_{m}({\tp}_{i})}\mathbb{E}e^{\delta t}V(x,t,\sigma_{t})\\
&\le& \frac{1}{\min_{i}\lambda_{m}({\tp}_{i})}V(x_0,0,\sigma_{0})
\end{eqnarray*}
So,
\begin{eqnarray*}
\mathbb{E}\|x^{j}(t)-s(t)\|^{2}\le V(y_0,0,\sigma_{0})e^{-\delta t}.
\end{eqnarray*}
The proof is completed.\hfill$\square$

As an application, we give the following theorem. Denote $q_i=-q_{ii}$.
\begin{theorem}\label{slowapp}
Suppose $\Gamma=I_m$, $f(x^i,t)\in QUAD(G,\alpha\Gamma,\beta)$ and every
$L(i)$ is strongly connected, then there exists diagonal positive definite
matrices $P_i$, coupling strength $\kappa$ and scale $\epsilon>0$, such
that $\{P_i[\alpha I_{m}+\kappa L(i)-\kappa\epsilon C(i)]\}^s$ are negative
definite. If
\begin{eqnarray}\label{app1}
q_{i}\leq -\frac{\lambda_{M}\{P_i[\alpha I_{m}+\kappa L(i)-\kappa\epsilon C(i)]\}^s}{\max_{j}\lambda_M(P_j)}
\end{eqnarray}
then the system (\ref{syst1}) is exponentially stable at the homogeneous trajectory in mean square sense.
\end{theorem}
{\em Proof.~} Let $[p_{1}^{i},\cdots,p_{m}^{i}]^{\top}$ be the left
eigenvector of matrix $L(i)$ corresponding to eigenvalue 0. \cite{Horn} has
proved that if $L(i)$ is strongly connected, all $p_j^i>0, i=1,\cdots,N,
j=1,\cdots,m$. Denote $P_{i}=diag\{p_{1}^{i},\cdots,p_{m}^{i}\}$. In
\cite{Chen07}, the authors proved that $\{P_i[\kappa L(i)-\kappa\epsilon
C(i)]\}^s, i=1,\cdots,N$ are negative definite. Therefore, we can find
suitable $\kappa>0,\epsilon>0$ such that $\{P_i[\alpha I_{m}+\kappa
L(i)-\kappa\epsilon C(i)]\}^s, i=1,\cdots,N$ are negative definite. In
addition, from the assumptions of $\sigma_t$, we know $q_{ii}=-q_i<0$ and
$\sum_{j\neq i}q_{ij}=q_i \sum_{j\neq i}p_{ij}=q_i$. Hence, if $q_i$
satisfies condition (\ref{app1}), we have
\begin{align*}
&\{P_i[\alpha I_{m}+\kappa L(i)-\kappa\epsilon C(i)]\}^s+\sum_{j=1}^N q_{ij}P_j\\
&\leq\lambda_{M}\{P_i[\alpha I_{m}+\kappa L(i)-\kappa\epsilon C(i)]\}^s
+q_i \max_{j}\lambda_M(P_j)\\
&\leq 0, ~~~~~~~~~~~~~~~~~~~~~~ i=1,\cdots,N.
\end{align*}
The rest is to apply theorem \ref{slow}. The proof is completed.
\begin{remark}
This theorem indicates that if each subsystem in the switching system can
be stabilized by the fixed pinning node set and the Markovian switching is sufficiently
slow, then the system is stable.
\end{remark}
\begin{remark}
In fact, if $\{P_i[\alpha I_{m}+\kappa L(i)-\kappa\epsilon C(i)]\}^s<0$
does not hold for some index $i$, with  properly picked $P_{j}$, $j\ne i$, for example, $P_{i}>P_{j}$, sufficiently large $-q_{ii}$, i.e, a small sojourn time at $i$,
properly picked $q_{ji}$
the condition (\ref{slow condition}) is possible to hold for $i$. Hence, Theorem 1 is more general than
Theorem 2 and the switching system among stable and unstable subsystems may be stable if both the sojourn time in these unstable
subsystems and transition probabilities from stable subsystems to them are sufficiently small.
\end{remark}

\section{Pinning fast switching networks}
In this section, we will demonstrate that if the Markov chain $\sigma_t$
has invariant distribution, its corresponding generator $Q$ and invariant
distribution $\pi$ play key roles in the stability analysis. The diagonal
elements of generator reflect the switching speed. The invariant
distribution is used to construct a network called average network. For the
Markov chain $\sigma_t$ with a unique invariant distribution
$\pi=[\pi_{1},\cdots,\pi_{N}]$, denote $\bar{L}=\sum_{i=1}^{N}\pi_{i}L(i)$
the average matrix of $L(\sigma_t)$ and $\bar{C}=\sum_{i=1}^{N}\pi_{i}C(i)$
the average matrix of $C(\sigma_t)$. For the fast switching case, we get
the pinning controllable criteria of (\ref{syst1}) from the average system.


Before giving the results, we restate the well-known Borel-Cantelli lemma
in the form presented in \cite{Kushner}.
\begin{lemma}\label{Borel}
For a stochastic process $X_k\in\mathbb{R}^n$ with $k\in \mathbb{Z}$ and a
nonnegative function $g: \mathbb{R}^n\rightarrow \mathbb{R}$, if
\begin{eqnarray*}
\sum_{k=0}^{\infty}\mathbb{E}[g(X_k)]<\infty,
\end{eqnarray*}
then $g(X_k)$ converges to zero almost surely.
\end{lemma}
\begin{theorem}\label{fast}
Suppose $f(x^i,t)$ is a Lipschitz function with Lipschitz constant $Lf$,
$f(x^i,t)\in QUAD(G,\alpha\Gamma,\beta)$
and there exist a diagonal
positive definite matrix $P$, constants $\kappa>0$ and $\epsilon>0$ such
that $\{P(\alpha I_{m}+\kappa\bar{L}-\kappa\epsilon\bar{C})\}^{s}$ is
negative semidefinite. If $\min_{i}q_i >\frac{1}{r\Delta}$ for some
$r\in\mathbb{N}$ and here $\Delta>0$ satisfies
\begin{eqnarray}\label{co}
\nonumber&&-\frac{(K_1-K_3)\lambda_{m}({\tp})}{\rho\lambda_{M}({\tp})}(1-e^{-\rho\Delta})\\
&&+\frac{K_4\lambda_{M}({\tp})\Delta}{\rho\lambda_{m}({\tp})}(e^{\rho\Delta}-1)< 0
\end{eqnarray}
where
\begin{align*}
&{\tp}=P\otimes G, ~K_1=\beta \min_i({p_{ii}}),\\
&K_2=\max_{i,j}|\lambda_{j}(\{P(\alpha I_m+\kappa L(i)-\kappa\epsilon C(i))\otimes G\Gamma\}^s)|,\\
&K_3=\lambda_{M}(\{P(\alpha I_m+\kappa\bar{L}-\kappa\epsilon\bar{C})\otimes G\Gamma\}^s),\\
&K_4=mn[\max_{i}\|A(i)\|_{\infty}(1+Lf^2)\\
&+\max_{i,j}\|A(i)^s(\kappa L(j)-\kappa\epsilon C(j)\otimes \Gamma)\|_{\infty}],\\
&\rho=\frac{2K_2}{\lambda_m({\tp})}+\frac{2K_1}{\lambda_M({\tp})}>0,\\
&A(\sigma_t)=\{P[\kappa L(\sigma_{t})-\kappa\bar
L-\kappa\epsilon C(\sigma_{t})+\kappa\epsilon\bar C]\otimes
G\Gamma\}^{s}
\end{align*}

then the system (\ref{syst1}) is stabilized almost surely.
\end{theorem}

{\em Proof.} Let $x=[{x^{1}}^{\top},\cdots,{x^{m}}^{\top}]^{\top}$,
$F(x)=[f(x^{1})^{\top},\cdots,f(x^{m})^{\top}]^{\top}$, $y=x-\hat{s}$ and
define $ V(t)=\frac{1}{2}y^{\top}{\tp}y$. Then the derivation of $V$ along system (\ref{syst1}) satisfies:
\begin{eqnarray*}
|\dot{V}(t)|&\leq&|y^{\top}{\tp}[F(x)-F(s)-\alpha (I_{m}\otimes\Gamma)y]|\\
\nonumber &+&|y^{\top}\{{\tp}[\alpha I_{m}+\kappa L(\sigma_{\tau})-\kappa\epsilon C(\sigma_{\tau})]\otimes
\Gamma\}^{s}y|\\ \label{de}
&\le&2\bigg[\frac{K_2}{\lambda_m({\tp})}+\frac{K_1}{\lambda_M({\tp})}\bigg]V(t)
\end{eqnarray*}
Denote $\rho=\frac{2K_2}{\lambda_m({\tp})}+\frac{2K_1}{\lambda_M({\tp})}$. Thus, for any $\tau>t$,
\begin{eqnarray}\label{rho}
V(t)e^{-\rho(\tau-t)}\leq V(\tau)\leq V(t)e^{\rho(\tau-t)}
\end{eqnarray}
Denote
\begin{align*}
\Phi(t,t+\Delta)=\int_{t}^{t+\Delta}[y(\tau)^{\top}A(\sigma_\tau)y(\tau)d\tau
-y(t)^{\top}A(\sigma_\tau)y(t)]d\tau
\end{align*}
Hence,
\begin{align*}
\nonumber&\Phi(t,t+\Delta)\\
\nonumber&=\int_{t}^{t+\Delta}\int_{t}^{\tau}\frac{d}{d\theta}[y(\theta)^{\top}A(\sigma_\tau)y(\theta)]d\theta
d\tau\\
\nonumber&=2\int_{t}^{t+\Delta}\int_{t}^{\tau}y(\theta)^{\top}A(\sigma_\tau)[F(x)-F(s)]d\theta
d\tau \\
\nonumber&+2\int_{t}^{t+\Delta}\int_{t}^{\tau}y(\theta)^{\top}\{A(\sigma_\tau)[kL(\sigma_{\theta})-k\epsilon
C(\sigma_{\theta})]\otimes \Gamma\}^{s}y(\theta)d\theta d\tau\\
\end{align*}
It is known that for any matrix $B=[b_{ij}]_{m\times n}$ and any vectors $x\in \mathbb{R}^m, y\in \mathbb{R}^n$,
\begin{eqnarray*}
x^{\top}By\leq \frac{\max(m,n)}{2}\max_{i,j}|b_{ij}|(x^{\top}x+y^{\top}y)
\end{eqnarray*}
holds. Therefore,
\begin{align*}
&y(\theta)^{\top}A(\sigma_\tau)[F(x)-F(s)]\\
&\leq\frac{mn}{2}\max_{i}\|A(i)\|_{\infty}[y(\theta)^{\top}y(\theta)+(F(x)-F(s))^{\top}(F(x)-F(s))]\\
&\leq\frac{mn}{2}\max_{i}\|A(i)\|_{\infty}[1+Lf^2]y(\theta)^{\top}y(\theta)
\end{align*}
and
\begin{align*}
&y(\theta)^{\top}\{A(\sigma_\tau)[kL(\sigma_{\theta})-k\epsilon
C(\sigma_{\theta})]\otimes \Gamma\}^{s}y(\theta)\\
&\leq\frac{mn}{2}\max_{i,j}\|A(i)^s(\kappa L(j)-\kappa\epsilon C(j)\otimes \Gamma)\|_{\infty}y(\theta)^{\top}y(\theta).
\end{align*}
Then, we have
\begin{align*}
&\Phi(t,t+\Delta)\leq\frac{2K_4}{\lambda_m({\tp})}\int_{t}^{t+\Delta}\int_{t}^{\tau}V(\theta)d\theta d\tau\\
\nonumber
&\leq\frac{2K_4}{\lambda_m({\tp})}\Delta\int_{t}^{t+\Delta}V(\theta)d\theta.
\end{align*}
Based on the above inequalities, we have
\begin{align*}
&V(t+\Delta)-V(t)=\int_{t}^{t+\Delta}\dot{V}(\tau)d\tau\\
&=\int_{t}^{t+\Delta}y^{\top}\tilde{P}[F(x)-F(s)-\alpha(I_m\otimes\Gamma)y]d\tau\\
&+\int_{t}^{t+\Delta}y^{\top}\{\tilde{P}[\alpha I_m+\kappa L(\sigma_\tau)-\kappa\epsilon C(\sigma_\tau)]\otimes\Gamma\}^syd\tau\\
&\le -\frac{2K_1}{\lambda_{M}({\tp})}\int_{t}^{t+\Delta}V(\tau)d\tau+\Phi(t,t+\Delta)\\
&+\int_{t}^{t+\Delta}y(\tau)^{\top}\{P[\alpha I_m+\kappa\bar L-\kappa\epsilon\bar C]\otimes G\Gamma\}^{s}y(\tau)d\tau\\
&+\int_{t}^{t+\Delta}y(t)^{\top}A(\sigma_{\tau})y(t)d\tau\\
&\leq-\frac{2K_1-2K_3}{\lambda_{M}({\tp})}\int_{t}^{t+\Delta}V(\tau)d\tau+\frac{2K_4}{\lambda_m({\tp})}\Delta\int_{t}^{t+\Delta}V(\theta)d\theta\\
&+\int_{t}^{t+\Delta}y(t)^{\top}A(\sigma_{\tau})y(t)d\tau
\end{align*}

Note that the conditional expectation
\begin{eqnarray*}
&&\mathbb{E}[\int_{t}^{t+\Delta}(L(\sigma_{\tau})-\bar{L})d\tau|y(t)]\\
&=&\sum_{k=k_0+1}^{k_0+n_k}
\sum_{j=1}^{N}L(j)\pi_j(k)\mathbb{E}(\Delta_{j}^{k})-
\bar{L}\Delta\\
&=&n_k\sum_{j=1}^{N}L(j)\pi_j\frac{1}{q_j}-\bar{L}\Delta+R_1(t)=R_1(t)
\end{eqnarray*}
where $k_0$ is the switching times before time $t$,
$n_{k}=\frac{\Delta}{\sum_{j=1}^{N}{\pi_j\frac{1}{q_j}}}$ is the expected
switching times between interval $[t,t+\Delta]$, $\Delta_{j}^{k}$ is the
sojourn time in state $L(j)$ and $R_1(t)$ is defined by:
\begin{eqnarray*}
R_1(t)=\sum_{k=k_0+1}^{k_0+n_k}\sum_{j=1}^{N}L(j)(\pi_j(k)-\pi_j)\frac{1}{q_j}
\end{eqnarray*}
Then,
\begin{eqnarray*}\label{R(t)}
\|R_1(t)\|_{\infty}&\leq& \sum_{k=k_0+1}^{k_0+n_k}\sum_{j=1}^{N}\|L(j)\|_{\infty}|\pi_{j}(k)-\pi_j|\frac{1}{q_j}\\
&\leq& \max_i\|L(i)\|_{\infty} M\zeta^{k_0}\sum_{j=1}^N\frac{n_k}{q_j}\leq M_1 \Delta\zeta^{k_0}
\end{eqnarray*}
where for any $k\in \mathbb{N}^+$ and $j=1,\cdots,N$, $M_1=M\max_i\|L(i)\|_{\infty}\sum_{j=1}^{N}\frac{1/{q_j}}{\sum_{j=1}^{N}{\pi_j/{q_j}}}$. Similarly, we can obtain $\mathbb{E}[\int_{t}^{t+\Delta}(C(\sigma_{\tau})-\bar{C})d\tau|y(t)]=R_2(t)$ with $\|R_2(t)\|_{\infty}\leq M_2\Delta\zeta^{k_0}$ with some $M_2>0$. Hence,
\begin{align}
\label{con-exp}&\mathbb{E}[V(t+\Delta)|y(t)]-V(t)\\
\nonumber&\leq[-\frac{2(K_1-K_3)}{\rho\lambda_{M}({\tp})}(1-e^{-\rho\Delta})+\frac{2K_4\Delta}{\rho\lambda_{m}({\tp})}(e^{\rho\Delta}-1)]V(t)\\
\nonumber&+y(t)^{\top}\{P(\kappa R_1(t)+\kappa\epsilon R_2(t))\otimes G\Gamma\}^s y(t)\\
\nonumber
\end{align}
the inequality is derived from (\ref{rho}). For any given $\Delta$ and $K_5>0$, we can find integer $\bar k(\Delta)$ such that
\begin{eqnarray}
mn\kappa\|P\|_{\infty}\|G\Gamma\|_{\infty}\zeta^{\bar k(\Delta)}(M_1+\epsilon M_2)<K_5
\end{eqnarray}
A reasonable requirement is that the average sojourn time $\sum_{j=1}^{N}\pi_j/q_j$ should be finite. We suppose there is an integer $r>0$ such that
\begin{eqnarray*}
\sum_{j=1}^{N}\frac{\pi_j}{q_j}\leq\max_{j}\frac{1}{q_j}<r\Delta,
\end{eqnarray*}
which means $q_i$ needs to satisfy:
\begin{eqnarray*}
\min_{i}q_i>\frac{1}{r\Delta}.
\end{eqnarray*}
Therefore, there must exist a finite time $t_{\bar{k}(\Delta)}$ such that $\sigma_t$ has switched $\bar{k}(\Delta)$ times before $t_{\bar{k}(\Delta)}$. Hence, for any $t>t_{\bar{k}(\Delta)}$,
\begin{align*}
&y(t)^{\top}\{P(\kappa R_1(t)+\kappa\epsilon R_2(t))\otimes G\Gamma\}^s y(t)\\
&\leq mn\kappa\|P\|_{\infty}\|G\Gamma\|_{\infty}(M_1+\epsilon M_2)\Delta\zeta^{\bar k(\Delta)}
y(t)^{\top}y(t)\\
&<K_5 \Delta y(t)^{\top}y(t)
\end{align*}
holds. Notice that $\Delta$ satisfies condition (\ref{co}), hence, when $K_5$ is sufficiently small, we can find a positive constant $\gamma$ such that
\begin{eqnarray}\label{co'}
\nonumber&&-\frac{(K_1-K_3)\lambda_{m}({\tp})}{\rho\lambda_{M}({\tp})}(1-e^{-\rho\Delta})\\
&&+\frac{K_4\Delta\lambda_{M}({\tp})}{\rho\lambda_{m}({\tp})}(e^{\rho\Delta}-1)+K_5\Delta\leq-\gamma.
\end{eqnarray}
Take expectations of (\ref{con-exp}), we obtain for any $t>t_{\bar{k}(\Delta)}$,
\begin{align}
\nonumber&\mathbb{E}[V(t+\Delta)]-\mathbb{E}[V(t)]\\ \nonumber
&\leq-\frac{(K_1-K_3)\lambda_{m}({\tp})}{\rho\lambda_{M}({\tp})}(1-e^{-\rho\Delta})
\mathbb{E}[y(t)^{\top}y(t)]\\ \nonumber
&+\frac{K_4\lambda_{M}({\tp})\Delta}{\rho\lambda_{m}({\tp})}(e^{\rho\Delta}-1)\mathbb{E}[y(t)^{\top}y(t)]\\
\nonumber&+K_5\Delta\mathbb{E}[y(t)^{\top}y(t)]\\
&\leq-\gamma\mathbb{E}[y(t)^{\top}y(t)]\label{expect}
\end{align}
Iterating (\ref{expect}), we obtain for any integer $p$,
\begin{eqnarray}\label{iterate}
\nonumber&&\mathbb{E}[V(t+p\Delta)]-\mathbb{E}[V(t)]\\
&\leq&-\gamma\sum_{i=1}^{p}\mathbb{E}[y(t+i\Delta)^{\top}y(t+i\Delta)].
\end{eqnarray}
Thus, (\ref{iterate}) means $\sum_{i=1}^{p}\mathbb{E}[y(t+i\Delta)^{\top}y(t+i\Delta)]$ is bounded for any $t>t_{\bar{k}(\Delta)}$ and any integer $p$. Apply lemma \ref{Borel} to $g(y)=y^{\top}y$, we obtain that $y(t)$ converges to zero almost surely.
\hfill$\square$

\begin{remark}
We have mentioned if the average coupling matrix $\bar{L}$ is strongly
connected, then there exist postive constants $\alpha,\kappa,\epsilon$ and a positive diagonal matrix $P$ such that $\{P(\alpha I_{m}+\kappa\bar{L}-\kappa\epsilon\bar{C})\}^{s}$ is negative
semidefinite. Condition (\ref{co}) can be guaranteed if $q_i$ is
sufficiently large. Hence, this theorem indicates that if the system with
the average coupling matrix and average pinning gains can be stabilized and
the switching is sufficiently fast, the time-varying system is stabilized.
\end{remark}

As an application, we consider spatial pinning control of network with
mobile agents. Inspired by \cite{Frasca}, we consider $m$ agents moving in
the planar space $\Lambda\subseteq \mathbb{R}^{2}$ according to the random
waypoint model \cite{Johnson}. Denote $y_{i}(t)=(y_{i1}(t),y_{i2}(t))\in
\Lambda$ as the position of agent $i$ at time $t$. The motion of each agent
is stochastically independent of the other ones but follows a identical
distribution. The agent moves towards a randomly selected target with a
randomly velocity. After approaching the target, the agent waits for a
random time length and then continues the process. According to the
positions of agents in the given area $\Lambda$, a undirected graph can be
structured. Two agents are considered to be linked at time $t$ if the
distance between them is less than a given interaction radius $r>0$.
Therefore, the coupling matrix is assigned by the spatial distribution of
the agents at each time. In detail, if $\|y_{i}(t)-y_{j}(t)\|\leq r$,
agents $i$ and $j$ are coupled with adjacent coefficient 1, otherwise, they
are uncoupled. Hence, the coupling matrix is a finite dimensional symmetric
matrix with elements equal to 0 or 1. Denote
$\mathbb{S}=\{L(1),\cdots,L(N_1)\}$ all possible topologies of the $m$
agents and $L(\sigma_t)$ the coupling matrix of the network at time $t$,
here $\sigma_t:\mathbb{R}^+\rightarrow\{1,\cdots,N_1\}$. We can consider
$L(\sigma_t)$ as a homogeneous continuous-time Markov chain with finite
state space $\mathbb{S}$.

The pinning control region $\Lambda_{c}$ is a fixed region in $\Lambda$
such that once agent $i$ enters the control region, i.e.,
$y_{i}(t)\in\Lambda_{c}$, then a control input is applied to agent $i$,
that is, agent $i$ is pinned. For a pinned subset $E$ of $\{\onetom\}$, we
define a pinning control matrix $C$ in such a way: if $i\in E$, then
$C_{ii}=1$, otherwise $C_{ii}=0$ and $C_{ij}=0$ for $i\neq j$. Denote
$\mathbb{C}=\{C(1),\cdots,C(N_2)\}$ all possible pinning control matrices
and $C(\eta_t)$ the pinning control matrix at time $t$, here
$\eta_t:\mathbb{R}^+\rightarrow\{1,\cdots,N_2\}$. Similarly, $C(\eta_t)$ is
also a homogeneous continuous Markov chain with a finite state space
$\mathbb{C}$. This effect is due to the motion of agents follow independent
Markov processes and the relation between the coupling matrix and the
position of agents.

For the random waypoint, there are two states: one is the moving state in
which the agent is moving towards a preselected target and the other is the
waiting state in which the agent is waiting for the next movement. Thus,
letting $V$ be the set of agents, we can consider the location and state of
the agents in the area as a stochastic process
$\chi^t=[(\alpha_i^t,\beta_i^t),v_i^t,w_i^t,\xi_i^t,(\eta_i^t,\vartheta_i^t)]_{i\in
V}$, where $(\alpha_i^t,\beta_i^t)$ denote the coordinates of agent $i$ at
time $t$; $v_i^t$ is the velocity of this movement and is zero if the agent
$i$ is waiting; $w_i^t$ is the current waiting time period if agent $i$ is
waiting and zero if it is moving; $\xi_i^t$ is the time cost of the current
wait if agent $i$ is waiting, and -1 if it is moving; and
$(\eta_i^t,\vartheta_i^t)$ are the coordinates of the target of the
movement of agent $i$ and equal to a fixed position which is out of the
region if it is waiting. Thus, $\chi^t$ is a homogeneous Markov chain.
Hence $\{\sigma_t,\eta_t\}$ is a higher dimensional homogeneous Markov
chain. For convenience, we suppose $\sigma_t=\eta_t$ in the following.

The node distribution of the RWP model was studied by
Bettstetter et.al. (2003). Their result implies that the
node distribution of RWP is ergodic and its stationary
pdf is always positive everywhere in the permitted region.
Each pair of agents has a positive probability to be linked.

Note that the motion of each agent is independent with each other and
follows a same distribution, which implies that the expected coupling
matrix $\bar{L}$ is complete and all of its non-diagonal elements are equal
and the diagonal elements of the expected  pinning matrix $\bar{C}$ are all
positive and equal.

On the other hand, it is easy to see that the switching of Markovian
processes $L(\sigma_t)$ will be fast if the mobile agents move sufficiently
fast and the waiting time period is sufficiently short. Denote $d$ the
maximum distance of each pair of plots in area $\Lambda$ and $\bar{p}$ the
probability that the distance between the targets of any two agent is more
than $r>0$. Denote $\underline{v}=\min_{i}v_i^t$ the minimum moving
velocity  and $\bar{t}_{w}=\max_{i}w_i^t$ the maximum waiting time.
Apparently, the moving time of an agent is less than
$\bar{t}_{m}=\frac{d}{\underline{v}}$.
Follow the mark mentioned above, $(\alpha_i^t,\beta_i^t)$ and
$(\theta_i^t,\vartheta_i^t)$ denote the coordinates of $i$ and its target
respectively. Suppose agent $i$ costs more time to achieve its target than
agent $j$. If the minimum waiting time is larger than $\bar{t}_{m}$ and the
distance between the targets of agents $i,j$ is larger than $r$, then
after time
$\frac{\|(\alpha_i^t,\beta_i^t)-(\theta_i^t,\vartheta_i^t)\|}{v_i^t}<\bar{t}_m$,
the distance between agent $i$ and $j$ is more than $r$. For two agents $i,j$ which are
coupled currently, denote $w_{ij}$ the expected time of agent $i$ escaping
from agent $j$'s $r-$ disc, i.e. to whose distance is less than $r$. It can be
estimated:
\begin{eqnarray*}
w_{ij}&<&\sum_{n=0}^{\infty}\bar{p}(1-\bar{p})^n((n+1)\bar{t}_m+n\bar{t}_{w})\\
&<&\frac{1}{\bar{p}}\bar{t}_m+\frac{(1-\bar{p})\bar{t}_w}{\bar{p}}=
\frac{d}{\bar{p}\underline{v}}+\frac{(1-\bar{p})\bar{t}_w}{\bar{p}}.
\end{eqnarray*}

Similarly, we can estimate the average escape time $e_i$ of agent $i$ to
get out of the pinning region:
$e_i<\frac{\tilde{d}}{\tilde{p}\underline{v}}+\frac{(1-\tilde{p})\bar{t}_w}{\tilde{p}}$,
where $\tilde{p}$ is the probability that the coordinate of the target of
agent $i$ is out of the pinning region and $\tilde{d}$ is the maximum
distance of $\Lambda_{c}$ Denote $1/{q_i}$ the expected sojourning time
period of $L(\sigma_t)=L(i)$. Apparently, they satisfy:
$\max_{i}\frac{1}{q_i}<\max(\max_{i,j}w_{ij},\max_i e_i)$. Hence, the
assumption of $\max(\max_{i,j}w_{ij},\max_i e_i)<r\Delta$ can derive
$\min_{i}q_i>\frac{1}{r\Delta}$.

Denote $\bar{L}=l(\bold{1}\cdot\bold{1}^{\top})-mlI_m$, here $l$ is the
probability that two agents are coupled. Denote $\bar{C}=cI_m$, here $c$ is
the probability that an agent enters into pinning region $\Lambda_c$.
Hence, the eigenvalues of $\{\alpha
I_m+\kappa\bar{L}-\kappa\epsilon\bar{C}\}^s$ are $\alpha-\kappa\epsilon c$
and $\alpha-\kappa ml-\kappa\epsilon c$. Then, for any $\alpha$, we can
find $\kappa,\epsilon>0$ such that $\alpha-\kappa\epsilon<0$, which implies
that $\{\alpha I_m+\kappa\bar{L}+\kappa\epsilon\bar{C}\}^s$ is negative
semi-definite. Apply theorem \ref{fast} to the above network, the following
corollary is obtained.

\begin{corollary}
Pick $\kappa,\epsilon>0$ such that $\alpha-\kappa\epsilon c<0$, here $c$ is the probability that an agent enters into pinning region $\Lambda_c$. If $\frac{d}{\bar{p}\underline{v}}+\frac{(1-\bar{p})\bar{t}_w}{\bar{p}}<r\Delta$ and $\frac{\tilde{d}}{\tilde{p}\underline{v}}+\frac{(1-\tilde{p})\bar{t}_w}{\tilde{p}}<r\Delta$ hold and here $r\in\mathbb{N},\Delta>0$ satisfies condition (\ref{co}), the mobile agent system (\ref{syst1}) is stabilized almost surely.
\end{corollary}

\section{Numerical simulations}
\begin{figure}[!t]
\begin{center}
\includegraphics[width=.45\textwidth]{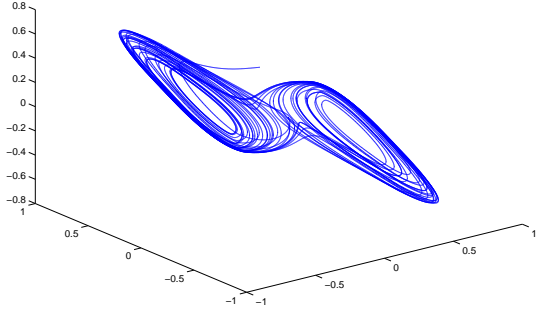}
\end{center}
\caption{Chaotic trajectory of system (\ref{neur}).} \label{f1}
\end{figure}
In this section, we give some numerical examples to illustrate the
theoretical results. In these examples, we consider three-dimensional
neural network as the uncoupled node dynamics \cite{Zou}:
\begin{equation}\label{neur}
\frac{dx}{dt}=-Dx+Tg(x)
\end{equation}
with $x=(x_1,x_2,x_3)^{\top}\in\mathbb{R}^3$,
\begin{eqnarray*}
T=\left[\begin{array}{ccc}
 1.2500 & -3.200 & -3.200 \\
  -3.200 & 1.100 & -4.400 \\
  -3.200 & 4.400 & 1.000
  \end{array}\right]
\end{eqnarray*}
$D=I_3$ and $g(x)=(g(x_1),g(x_2),g(x_3))^{\top}$ where
$g(s)=(|s+1|-|s-1|)/2$. This system has a double-scrolling chaotic
attractor shown in Fig.\ref{f1} with initial condition
$x_1(0)=x_2(0)=x_3(0)=0.1000.$ In the following, we suppose system
(\ref{neur}) is the uncoupled dynamical system of each agents and the inner
coupling matrix $\Gamma=I_n$. We can find that any $\alpha>0.5$ can satisfy
the decreasing condition with $G=I_{n}$. We use the following quantity to
measure the variance for vertices and target trajectory:
\begin{eqnarray*}
\varsigma(t)=\max_{i}\|x^{i}(t)-s(t)\|_1
\end{eqnarray*}
here for a vector $z=[z_1,\cdots,z_n]^{\top}\in\mathbb{R}^n$, denote $\|z\|_1\hat{=}\max_{i}|z_i|$.

\subsection{Slow switching among stable sub-systems}

\begin{figure}[!t]
\begin{center}
\includegraphics[width=.45\textwidth]{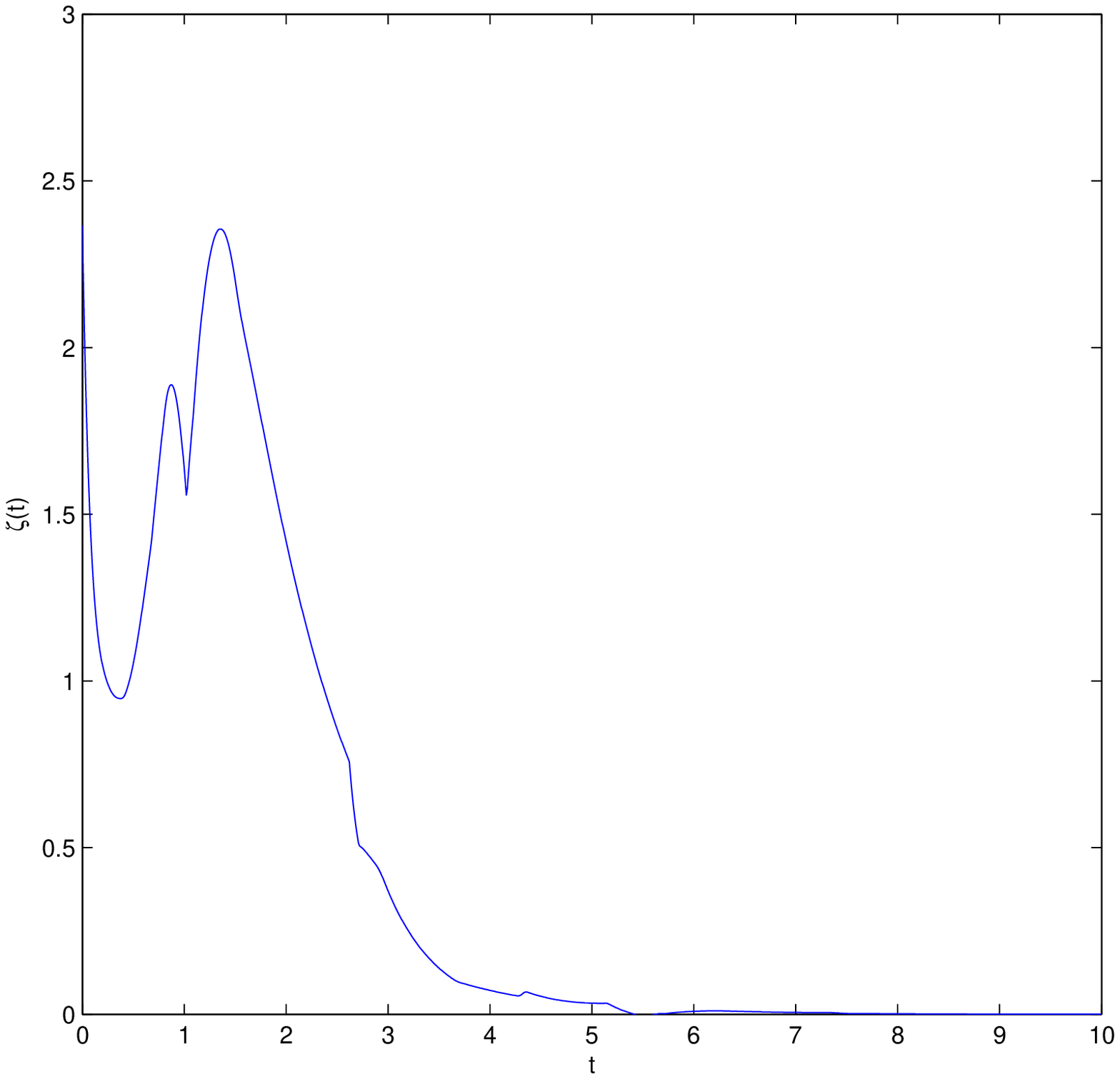}
\end{center}
\caption{The dynamical behavior of $\varsigma(t)$.} \label{fig1}
\end{figure}

This simulation is for the pinning controlled network (\ref{syst1}) under
assumption in theorem \ref{slowapp} with $m=5$. The transition matrix $T$
of the Markov chain $\sigma_{t}$ is
\begin{eqnarray*}
T=\left[
\begin{array}{ccccc}
  0 & 0.65 & 0 & 0.35 & 0 \\
  0 & 0 & 0.7 & 0 & 0.3 \\
  0 & 0.1 & 0 & 0.9 & 0 \\
  0.4 & 0.6 & 0 & 0 & 0 \\
  0 & 0.3 & 0 & 0.7 & 0
\end{array}
\right].
\end{eqnarray*}
Pick the coupling matrices
$$L(1)=
\left[\begin{array}{ccccc}
-3   &  0   &  1 &   1   &  1\\
     0   & -2  &   0  &   1  &   1\\
     1   &  1  &  -3   &  0   &  1\\
     0   &  0  &   0   & -1   &  1\\
     1   &  1  &   0   &  0   & -2
\end{array}\right]$$
$$L(2)=
\left[\begin{array}{ccccc}
 -2   &  0   &  0  &   1  &   1\\
     1   & -2    & 0    & 0 &    1\\
     0   &  1   & -2   &  0&     1\\
     0   &  1    & 0   & -2 &    1\\
     0  &   0   &  0    & 1 &   -1
\end{array}\right]$$
$$L(3)=
\left[\begin{array}{ccccc}
 -3  &   0  &   1  &   1   &  1\\
     1  &  -1  &   0 &    0&     0\\
     1   &  0   & -2 &    0 &    1\\
     0  &   0  &  0  &  -1  &   1\\
     1   &  1&     1  &   1  &  -4
\end{array}\right]$$
$$L(4)=
\left[\begin{array}{ccccc}
 -2   &  1   &  0   &  0 &    1\\
     1  &  -2  &   0   &  0&  1\\
     0   &  1  &  -3   &  1 &    1\\
     0   &  1  &   1  &  -2  &   0\\
     1   &  0  &   0  &   1   & -2
\end{array}\right]$$
$$L(5)=
\left[\begin{array}{ccccc}
 -2   &  1   & 1  &   0   &  0\\
     0  &  -3  &   1 &    1    & 1\\
     1  &   1  &  -4  &   1   &  1\\
     0  &   1  &   1  &  -3  &   1\\
     0  &   1 &    1  &   1 &   -3
\end{array}\right]$$
and the pinning control matrices
$$C(1)=diag\{1,1,0,0,1\}$$
$$C(2)=diag\{1,1,1,1,1\}$$
$$C(3)=diag\{0,0,0,1,1\}$$
$$C(4)=diag\{0,1,0,1,0\}$$
$$C(5)=diag\{1,1,1,0,0\}$$
Pick $P_i=I_m, \alpha=1, \beta=0.5$, $k=10$ and $\epsilon=1$, then
$\{\alpha I_{m}+kL(i)-k\epsilon C(i)\}^s$ are negative definite and
$\{\alpha I_{m}+kL(i)-k\epsilon C(i)\}^s\leq -0.75I_m$. In case
$\max_{i}{q_{i}}<0.75$, the condition in Theorem \ref{slow} is satisfied.
Choose $q_{i}$ randomly in $(0,0.75)$. The initial value $x(0)$ and $s(0)$
are also chosen randomly. The ordinary different equations (\ref{syst1})
are solved by the Runge-Kutta fourth-order formula with a step length of
0.01. Fig.\ref{fig1} indicates that the pinning control of (\ref{syst1}) is
stable.

\subsection{Locally pinning a system of mobile agents}
\begin{figure}[!t]
\begin{center}
\includegraphics[width=.45\textwidth]{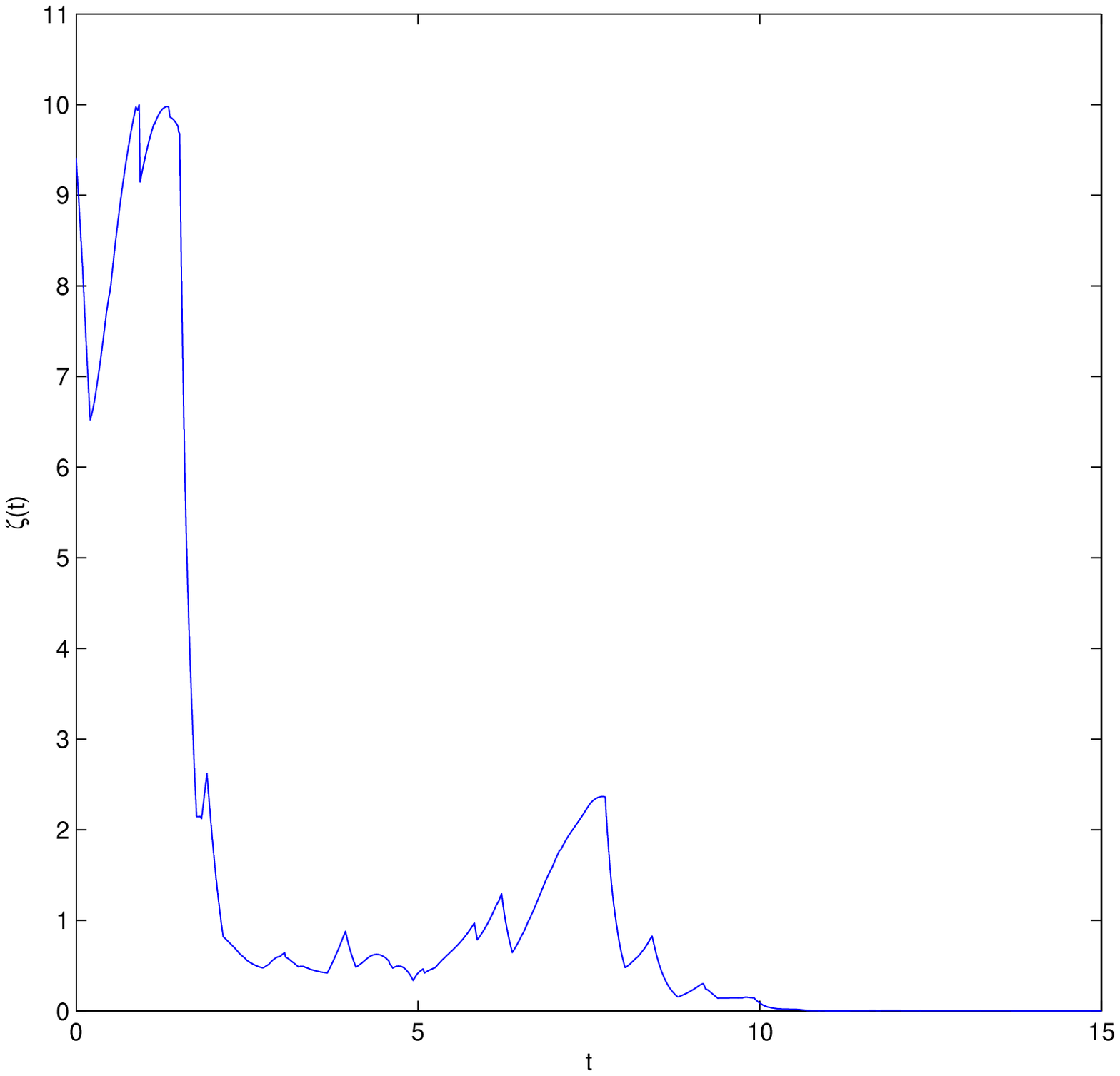}
\end{center}
\caption{The dynamical behavior of $\varsigma(t)$.} \label{fig2}
\end{figure}

In this simulation, we take the random waypoint mobility model
\cite{Bettstetter} with $10$ agents. Suppose  random waypoint area is
square $\Lambda=\{(y_{1},y_{2})\in \mathbb{R}^2:0\leq y_{1}\leq 100, 0\leq
y_{2}\leq 100\}$, and the agents are randomly located in $\Lambda$, the
control region $\Lambda_{c}=\{(y_{1},y_{2})\in \mathbb{R}^2:0\leq y_{1}\leq
50, 0\leq y_{2}\leq 50\}$. The initial location of each agent is randomly
chosen in $\Lambda$. If an agent achieves its target, it will stay in this
location during a randomly chosen time interval $a$. If the distance
between locations of two agents is less than 10, then these two agents are
coupled. If the location of an agent is in $\Lambda_{c}$ at time $t$, then
this agent is pinned at time $t$. Here, we pick $\alpha-\beta=0.5, P=I_m$.
In this simulation, we can calculate the expected matrices
$\bar{L}=[\bar{l}_{ij}]$ with $\bar{l}_{ij}=\frac{\pi 10^2}{100^2}\approx 0.0314$ for $i\neq j$,
 $\bar{C}=0.25I_m$. Pick $\kappa\epsilon>0.5$ can guarantee $\alpha-0.25\kappa\epsilon\leq 0$. It can be calculated that $K_1=\beta,K_2\leq\alpha, K_3=\alpha-0.25\kappa\epsilon$. The Lipschitz constant of function $-Dx+Tg(x)$ is $Lf=4.68$. Pick $\kappa=0.5,\epsilon=12$, we can estimate that $K_1-K_3=1, K_4=2500,\rho=33$. If we take $\Delta=0.0004$, the condition (\ref{co}) is guaranteed. Pick $r=750$. Hence, the minimum moving velocity $\underline{v}$ and maximum waiting time $\bar{t}_w$ should satisfy
$\frac{d}{\bar{p}\underline{v}}+\frac{(1-\bar{p})\bar{t}_w}{\bar{p}}<r\Delta$ and $\frac{\tilde{d}}{\tilde{p}\underline{v}}+\frac{(1-\tilde{p})\bar{t}_w}{\tilde{p}}<r\Delta$, here the probabilities can be estimated: $\bar{p}>1-\frac{\pi r_0^2}{100^2}>0.99$ and $\tilde{p}=\frac{3}{4}$. Then, we obtain that $\underline{v}>500$ and $\bar{t}_w<0.33$. On the other hand, the maximum moving time is $\bar{t}_m=\frac{d}{\underline{v}}<0.29$ and we assume the waiting time larger than 0.29. The ordinary different equations (\ref{syst1}) are solved by the Runge-Kutta fourth-order formula with a step length of 0.0001. Fig.\ref{fig2} indicates that the pinning control of (\ref{syst1}) is stable.

\section{Conclusion}
In this paper, we investigate the pinning control problem of coupled dynamical systems with Markovian switching couplings and Markovian switching controller-set. We derive analytical conditions for the stability at the homogeneous trajectory of the uncoupled system. First, if each subsystem in the switching system with fixed pinning nodes can be stabilized and the Markovian switching is sufficiently slow, the system van be stabilized. Second, the switching system can be stabilized if the system with average coupling matrix and pinning gains can be stabilized and the Markovian switching is sufficiently fast. As an application, we also study the spatial pinning control problem of network with mobile agents. The effectiveness of the proposed theoretical results are demonstrated by two numerical examples.

\end{document}